\RequirePackage{tikz}

\documentclass[pdflatex, sn-vancouver]{sn-jnl}

\usepackage[nolist]{acronym}
\usepackage{subcaption}
\usepackage{bm}
\usepackage[normalem]{ulem}
\usepackage{graphicx}

\usepackage{tikz}
\usepackage{pifont}

\usetikzlibrary{automata, arrows.meta, positioning, decorations, calligraphy, decorations.pathreplacing, calc, backgrounds}

\usepackage{todonotes}
\usepackage{menukeys}

\definecolor{col1}{HTML}{0072BD}
\definecolor{col2}{HTML}{D95319}
\definecolor{col3}{HTML}{EDB120}
\definecolor{col4}{HTML}{7E2F8E}
\definecolor{col5}{HTML}{77AC30}
\definecolor{col6}{HTML}{4DBEEE}
\definecolor{col7}{HTML}{A2142F}
\definecolor{edgered}{HTML}{d42020}

\jyear{2023}%

\begin{document}

\begin{acronym}
\acro{cnn}[CNN]{Convolutional Neural Network}
\acro{lraspp}[Lite R-ASPP]{Lite Reduced Atrous Spatial Pyramid Pooling}
\acro{resnet}[ResNet]{Residual Network}
\acro{relu}[ReLU]{Rectified Linear Unit}
\acro{cam}[CAM]{Class Activation Map}

\acro{hmm}[HMM]{Hidden Markov Model}
\acro{nll}[NLL]{Negative Log Likelihood}
\acro{ece}[ECE]{Estimated Calibration Error}
\acro{map}[MAP]{Maximum A Posteriori}

\acro{vb}[VB]{virtual bronchoscopy}
\acro{pb}[PB]{phantom bronchoscopy}
\acro{rb}[RB]{real bronchoscopy}

\acro{icu}[ICU]{Intensive Care Unit}
\acro{emt}[EMT]{Electromagnetic Tracking}
\acro{slam}[SLAM]{Simultaneous Localization And Mapping}
\acro{sfm}[SfM]{Structure from Motion}
\acro{mser}[MSER]{Maximally Stable Extremal Regions}
\acro{svm}[SVM]{Support Vector Machine}
\end{acronym}

\title[ ]{BronchoGAN: Anatomically consistent and domain-agnostic image-to-image translation for video bronchoscopy}

\author*[1]{\fnm{Ahmad} \sur{Soliman}}

\author*[1]{\fnm{Ron} \sur{Keuth}}



\author[2]{\fnm{Marian} \sur{Himstedt}}\email{marian.himstedt@th-luebeck.de}

\affil*[1]{\orgdiv{Medical Informatics}, \orgname{University of L\"ubeck}, \orgaddress{\street{Ratzeburger Allee 160}, \postcode{23562} \city{L\"ubeck}, \country{Germany}}}

\affil*[2]{\orgdiv{Faculty of Electrical Engineering and Computer Science}, \orgname{University of Technology L\"ubeck}, \orgaddress{\street{Mönkhofer Weg 239}, \postcode{23909} \city{L\"ubeck}, \country{Germany}}}




\abstract{\textbf{Purpose:} The limited availability of bronchoscopy images makes image synthesis particularly interesting for training deep learning models. Robust image translation across different domains - virtual bronchoscopy, phantom as well as in-vivo and ex-vivo image data - is pivotal for clinical applications.  

\textbf{Methods:} This paper proposes BronchoGAN introducing anatomical constraints for image-to-image translation being integrated into a conditional GAN. In particular, we force bronchial orifices to match across input and output images. We further propose to use foundation model-generated depth images as intermediate representation ensuring robustness across a variety of input domains establishing models with substantially less reliance on individual training datasets. Moreover our intermediate depth image representation allows to easily construct paired image data for training. 

\textbf{Results:} Our experiments showed that input images from different domains (e.g. virtual bronchoscopy, phantoms) can be successfully translated to images mimicking realistic human airway appearance. We demonstrated that anatomical settings (i.e. bronchial orifices) can be robustly preserved with our approach which is shown qualitatively and quantitatively by means of improved FID, SSIM and dice coefficients scores. Our anatomical constraints enabled an improvement in the Dice coefficient of up to $0.43$ for synthetic images.
 
\textbf{Conclusion:} Through foundation models for intermediate depth representations, bronchial orifice segmentation integrated as anatomical constraints into conditional GANs  we are able to robustly translate images from different bronchoscopy input domains. BronchoGAN allows to incorporate public CT scan data (virtual bronchoscopy) in order to generate large-scale bronchoscopy image datasets with realistic appearance. BronchoGAN enables to bridge the gap of missing public bronchoscopy images. 
}


\maketitle

\section{Introduction}\label{sec:intro}

Bronchoscopy is a vital procedure frequently performed in pulmonology clinics, with the largest number of cases involving the examination and biopsy of patients suspected of having lung cancer. Beyond this primary application, bronchoscopy is also used for various other medical indications, such as monitoring patients with chronic obstructive pulmonary disease (COPD), performing biopsies in pneumonia cases, and addressing acute respiratory problems in intensive care unit (ICU) settings. 
While traditional computer vision has been integrated into several certified bronchoscopy-based intervention systems, the application of deep learning-based approaches holds immense potential to enhance these existing systems and enable novel approaches, particularly for navigation assistance as already motivated in \cite{keuth2024airway, sganga_autonomous_2019}. An adequate integration of deep learning-based models in medical products requires superior generalization capabilities which can only be obtained from a large patient cohort covering variability in anatomy and optical appearance. This is currently hampered by the lack of publicly available bronchoscopy images which are rarely recorded in clinical routine. Fortunately, large-scale public thorax CT scan datasets \cite{national2011NLST} comprising tens of thousands of volumes are available, offering a valuable resource. In addition, phantoms are frequently utilized in research and training \cite{visentini-scarzanella_deep_2017}, enabling the simulation of varying camera poses, motion dynamics, and lightning conditions. 
\textbf{This paper attempts to close the aforementioned gap by means of image-to-image translation rendering realistic bronchoscopy images from virtual bronchoscopy (VB) constructed from CT scans and phantom images.} 
Medical image translation has been investigated in-depth in recent years. Conditional GANs (cGAN) have been used for e.g. translating PET and CT \cite{zhou2023gan}, retinal images \cite{iqbal2018generative}, chest radiographs \cite{popp2024adaptive} and PET to MRI data \cite{chen2024ambient}. Wang et al. propose pix2pix \cite{wang2018pix2pixhd} using paired image data while cycleGAN \cite{zhu2017unpaired} and CUT \cite{park2020contrastive} can handle unpaired image sets. In \cite{atli2024i2i} authors propose an adversarial model for multi-modal synthesis integrating channel-mixed Mamba blocks into a CNN backbone, using selective state space modeling. More recently, denoising diffusion models (DDM) have been investigated for image translation promising more stable training and mode coverage compared to GANs at the cost of increased computational costs \cite{kazerouni2023diffusion, arslan2024self}. In \cite{arslan2024self}, authors introduce SelfRDB which further improves source-to-target alignment using a novel forward process, adaptive noise scheduling, and recursive sampling for enhanced accuracy. Exhaustive surveys are provided by \cite{zhou2023gan, dayarathna2024deep}. 
Image translation has also been used in conjunction with bronchoscopy. Existing approaches \cite{banach2021,xu2024depth,yang2024adversarial,ozbey2023unsupervised} primarily focus on GAN-based depth estimation motivated as pivotal fundamental for vision-only guidance. Only \cite{banach2021} proposes 3cGAN translating images between multiple input domains - virtual bronchoscopy, real bronchoscopy (RB) and depth - using individual generators and discriminators for each domain. However, also 3cGAN focuses on synthesizing depth images and rather incorporates VB data to ease ground-truth depth image acquisition.  
To our best knowledge, current state of the art \cite{banach2021,xu2024depth,yang2024adversarial,ozbey2023unsupervised} in bronchoscopy image translation misses two fundamental requirements:
\begin{enumerate}
    \item \textit{Anatomical consistency: }None of the existing approaches incorporates anatomical constraints, in particular bronchial orifices, in image translation even though the preservation of anatomical structures is indispensable for clinical applications.
    \item \textit{Domain-robustness: }Existing approaches are limited to specific input image domains with limited capabilities to generalize from unseen data (e.g. visual appearance across VB, phantom and RB images)
\end{enumerate}
While segmentation priors have already been integrated in other medical applications (e.g. \cite{hamghalam2024medical}), it is missing in state-of-the-art approaches to bronchoscopy image translation. Leveraging VB images from large-scale CT scan datasets \cite{national2011NLST} presents significant opportunities for deep learning-based models, particularly in accommodating the wide variety of lung anatomies incorporated by bronchial orifice segmentation \cite{keuth2023weakly}. 
The challenge of \textit{domain-robustness} is omnipresent in medical imaging particularly for applications with limited readily-available data. Substantial investigations have been done in the field of domain adaptation and generalization \cite{guan2021domain} as well domain randomization \cite{dinkar2022automatic}. Another option is the introduction of an intermediate image representation mitigating individual input domain variations and thus enabling \textit{domain-agnostic} image translation. A first approach to this has been presented in \cite{visentini-scarzanella_deep_2017}. Here, the authors propose to translate \textit{domain-specific} RB images to \textit{domain-agnostic} VB images in order to synthesize depth images \cite{visentini-scarzanella_deep_2017}. We take this idea further by using depth images as intermediate representations to synthesize real from virtual bronchoscopy images while also addressing a fundamental limitation of \cite{visentini-scarzanella_deep_2017}: the ability to achieve this intermediate representation as \textit{zero-shot }inference. 

Deep learning-based depth estimation has been extensively studied in numerous works (survey: \cite{ming2021deep}). Self-supervised approaches \cite{godard2019digging} have demonstrated superior performance compared to supervised methods, while a significant breakthrough for \textit{zero-shot} inference has been achieved with the advent of foundation models such as DINO \cite{oquab2023dinov2} and DepthAnything \cite{yang2024depth}. Depth images have found applications in several endoscopy-related tasks \cite{cui2024surgical}, including bronchoscopy \cite{guo2024cgan}, since they provide a \textit{domain-agnostic} representation, where the scene semantic is preserved. However, these applications have primarily motivated on leveraging depth data for direct benefits in 3D-reconstruction as well as visual mapping and tracking, except for \cite{wang_bronchial_2021,xu2024depth}. In this work, we adopt depth images as an intermediate, \textit{domain-agnostic} representation to address challenges associated with limited training datasets. This approach aims to mitigate adverse impacts on image translation, enabling more robust and generalized solutions.

\noindent \textbf{Key contributions:}
\begin{enumerate}
    \item The integration of a foundation model for estimating depth images as \textit{domain-agnostic} (intermediate) image representation 
    \item An extended cGAN \cite{wang2018pix2pixhd} model incorporating segmentation priors for anatomically consistent image translation with the goal of mimicking in-vivo and ex-vivo appearance
\end{enumerate}
Both, depth estimation \cite{yang2024depth} and bronchial orifice segmentation incorporated in our paper \cite{keuth_weakly_2022}, are \textit{training-free} pipelines making our approach particularly well-suited for \textit{domain-agnostic} image translation with limited data availability.

\section{Methods}\label{sec:methods}

\begin{figure*}
    \centering
    \includegraphics[width=.99\textwidth]{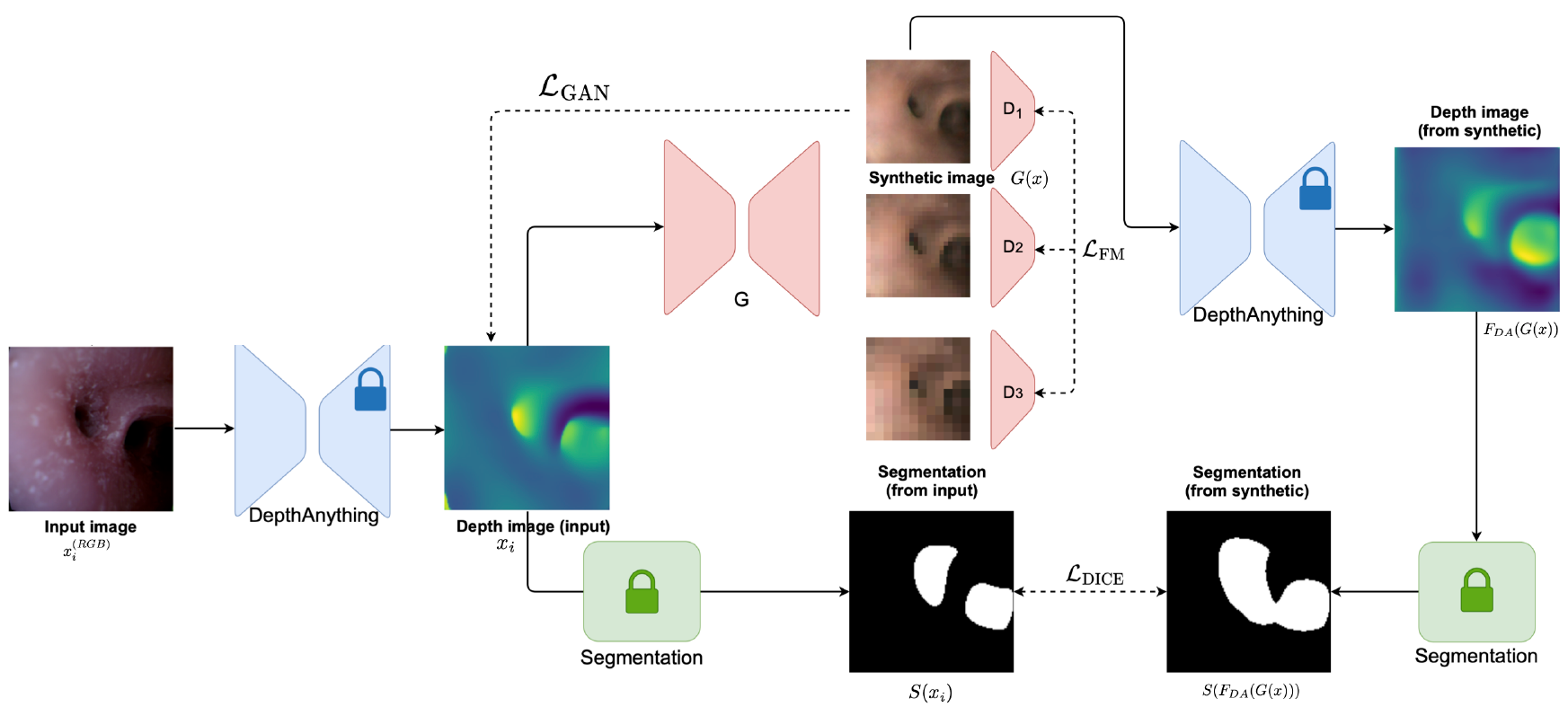}
    \caption{\textbf{BronchoGAN architecture}: RGB input images from virtual bronchoscopy and phantom datasets are processed by depthAnything generating a depth image as intermediate representation. A cGAN is trained on this depth images synthesizing real bronchoscopy using a hierarchical pix2pixHD. The output is translated to a depth image again. Bronchial orifices are segmented from both, input and output depth images using \cite{keuth2023weakly}. }
    \label{fig:architecture}
\end{figure*}


\subsection*{Overview} \label{sec:overview}
Our approach consists of multiple steps in order to prepare input data, estimate anatomical priors and finally translate images. In particular, we:
\begin{enumerate}
    \item Read an input bronchoscopy image (any bronchoscopy domain)
    \item Infer depth image from the input using a foundation model
    \item Segment bronchial orifices from (input) depth image and (output) synthetic image
    \item Translate depth image to synthetic in-vivo image constrained by forcing maximal overlap of input and output segmentations
\end{enumerate}
The following sections describe all steps, Fig. \ref{fig:architecture} visualizes the overall architecture.  

Initially, we are given a pair of corresponding images $\{x_i^{(RGB)}, y_i\}$, with $x_i^{(RGB)}$ denoting an input RGB image from a source domain and $y_i$ the corresponding RGB image in the target domain. 

\subsection*{Depth image estimation} \label{sec:depth}
We incorporate DepthAnything, a foundation model for depth inference from single monocular images. It is trained on a large dataset comprising 1.5M labeled and 62M unlabeled images. By utilizing pseudo-labels generated by a teacher model, it achieves zero-shot depth estimation, enabling generalization to unseen domains without additional training or fine-tuning. We utilize DepthAnything-V2 which generates depth images at higher quality by replacing labeled real images with synthetic data, improves pseudo-label generation through a more capable teacher model, and integrates extensive pseudo-labeled real images. 
Transforming bronchoscopy images into the depth representation enables more robust translation into our target domain. We denote $F_{DA}(x_i^{(RGB)})$ as the (inferred) output of DepthAnything generating a depth image $x_i$ from the input image $x_i^{(RGB)}$. Note that our approach uses depth as input for a GAN. Thus, we will refer to $x_i$ as a depth image.

\subsection*{Bronchial orifice segmentation} \label{sec:airway_seg}
A crucial requirement for medical image translation is the maintenance of anatomical properties to ensure the accuracy and reliability of the resulting images for clinical and diagnostic purposes. In this context, bronchial orifices, which serve as key anatomical landmarks, are segmented to preserve their structural integrity throughout the translation process. To achieve this, we employ a training-free pipeline designed to remain independent of the source domain. This independence ensures broad applicability across varying bronchoscopy image domains omitting individual training or fine-tuning. The segmentation is given a depth image which is searched for local extrema which refer to central locations of bronchial orifices. A subsequent non-maximum suppression rejects noisy peaks assuming a minimum distance to adjacent ones. Next, a k-means clustering is initialized at the peaks. The algorithm  differentiates bronchial orifices and other tissue. This enables precise separation of anatomical features, as detailed in our prior work \cite{keuth_weakly_2022}. The outcome of this process is a well-defined binary segmentation map delineating bronchial orifices from other tissue. A segmentation map $S(x_i)$ is generated using the segmentation model $S$ given an input depth image $x_i$.

\subsection*{Conditional GAN with anatomical constraints} \label{sec:orifice_det}
For image-to-image translation we use pix2pixHD \cite{wang2018pix2pixhd} as base architecture  comprising one generator $G$ and three discriminators $D_1$, $D_2$, $D_3$ working at different scales. The vanilla min-max optimization is given as:
\begin{equation}
\min_{G} \max_{D_1, D_2, D_3} \sum_{k=1,2,3} \mathcal{L}_{\text{GAN}}(G, D_k).
\end{equation}
The GAN loss $\mathcal{L}_{\text{GAN}}$ is stated as follows:
\begin{equation}
    \mathcal{L}_{\text{GAN}}(G, D) = \mathbb{E}_{(x, y)} \left[ \log D(x, y) \right] + \mathbb{E}_{x} \left[ \log \left( 1 - D(x, G(x)) \right) \right].
\end{equation}

The original model further includes a feature matching loss $\mathcal{L}_{\text{FM}}(G, D_k)$ comparing features from multiple layers of a discriminator to stabilize 
training and improve reconstruction quality (similarly to perceptual loss for VAE) \cite{wang2018pix2pixhd}. The feature extractor at layer $m$ of descriminator $D_k$ is referred to as $D^{m}_k$ with $T$ being the number of layers and $N_m$ the amount of elements in the corresponding layer:
\begin{equation}
\mathcal{L}_{\text{FM}}(G, D_k) = \mathbb{E}_{(\mathbf{x}, \mathbf{y})} \sum_{m=1}^{T} \frac{1}{N_m} \lVert D_k^{(m)}(\mathbf{x}, \mathbf{y}) - D_k^{(m)}(\mathbf{x}, G(\mathbf{x})) \rVert_1,
\end{equation}
We propose to include anatomical constraints in image translation by forcing maximal overlap of the segmentations extracted from the input depth image $S(x)$ and the depth-transformed output image constructed by the GAN $S(F_{DA}(G(x)))$. Our dice loss $\mathcal{L}_{\text{DICE}}$ penalizes image generation with few overlap of bronchial orifice segmentations ($\epsilon$ ensures numerical stability):
\begin{equation}
\mathcal{L}_{\text{DICE}}(G, S) = 1 - \frac{2 \sum_{j} S(x)^{(j)} S(F_{DA}(G(x)))^{(j)}}{\sum_{j} S(x)^{(j)} + \sum_{j} S(F_{DA}(G(x)))^{(j)} + \epsilon}
\end{equation}
Note that the index here iterates over single pixels $j$ and the foundation model's weights are fixed ($F_{DA}$).
Finally, we obtain the full objective function including anatomical constraints with $\mathcal{L}_{\text{FM}}$ and $\mathcal{L}_{\text{DICE}}$ being weighted by $\lambda_{\text{FM}}$ and $\lambda_{\text{DICE}}$:

\begin{equation}
\min_{G} \Bigg( \max_{D_1, D_2, D_3} \sum_{k=1,2,3} \big( \mathcal{L}_{\text{GAN}}(G, D_k) 
+ \lambda_{\text{FM}} \mathcal{L}_{\text{FM}}(G, D_k) \big) 
+ \lambda_{\text{DICE}} \mathcal{L}_{\text{DICE}}(G, S) \Bigg)
\end{equation}

\section{Experiments}\label{sec:results}
 
\subsection*{Setups} \label{sec:setups}
For training and testing we used a common workstation equipped with an Intel Core i7 with 32GB main memory and an NVIDIA A4000 with 16GB memory. We evaluated our proposed approach as follows: 

\noindent \textbf{pix2pix\_base}: A baseline architecture \cite{wang2018pix2pixhd}) was trained on paired \ac{vb} and \ac{pb} images from an airway phantom \cite{visentini-scarzanella_deep_2017} ($12\,000$ image pairs). Here, RGB images (\ac{vb}) served as input. This setup was limited since only \ac{pb} images from the phantom data could be used as pix2pixHD requires paired data. Also, we had to manually add circular crops to test images since the paired data published in \cite{visentini-scarzanella_deep_2017} contains those. 

\textbf{\noindent Note: pix2pix\_base could not be trained on other RB datasets since it is impossible to establish paired image sets. As already stated, this is the main benefit of using an intermediate representation (e.g. depth) for image translation as paired image data can be established at ease.}  

\noindent \textbf{pix2pix\_depth (ours)}: This setup extended the baseline by incorporating depthAnything to generate depth images as input for the GAN. The model was trained on all images of 2 public datasets - Edinburgh \cite{deng2023feature} and BI2K \cite{vu2024BI2K} - and 4 private datasets totaling to $12\,580$ image pairs. RB images from these datasets were processed by depthAnything to generate paired input depth and target RB images. Circular crops were not required for this setup. 

\noindent \textbf{BronchoGAN} (ours): This setup extended pix2pix\_depth by using bronchial orifice segmentation serving as anatomical constraints using $\mathcal{L}_{\text{DICE}}$. It was trained with the same data as pix2pix\_depth.

\noindent \textbf{cycleGAN}: A further baseline architecture \cite{zhu2017unpaired}) was trained on unpaired images. Here, we incorporated the same training data as pix2pix\_depth and BronchoGAN for target images. Instead of (DepthAnything-based depth) input data we used a total of $12\,580$ images with $1\,428$ phantom images \cite{visentini-scarzanella_deep_2017} and $11\,152$ VB images obtained from 20 unique patients \cite{national2011NLST}.  

All models were evaluated on the same test dataset, Harvard \cite{banach2022_harvard_data}, which consists of $2\,271$ VB images. The dice coefficients were estimated based on bronchial orifice segmentation maps constructed from input and synthesized images respectively using our approach presented in \cite{keuth_weakly_2022}. Note that our segmentation approach \cite{keuth_weakly_2022} working on depth images inferred from DepthAnything works training-free and has not been adapted for neither training nor test data.

\begin{figure*}[h]
    \centering
    \includegraphics[width=.99\textwidth]{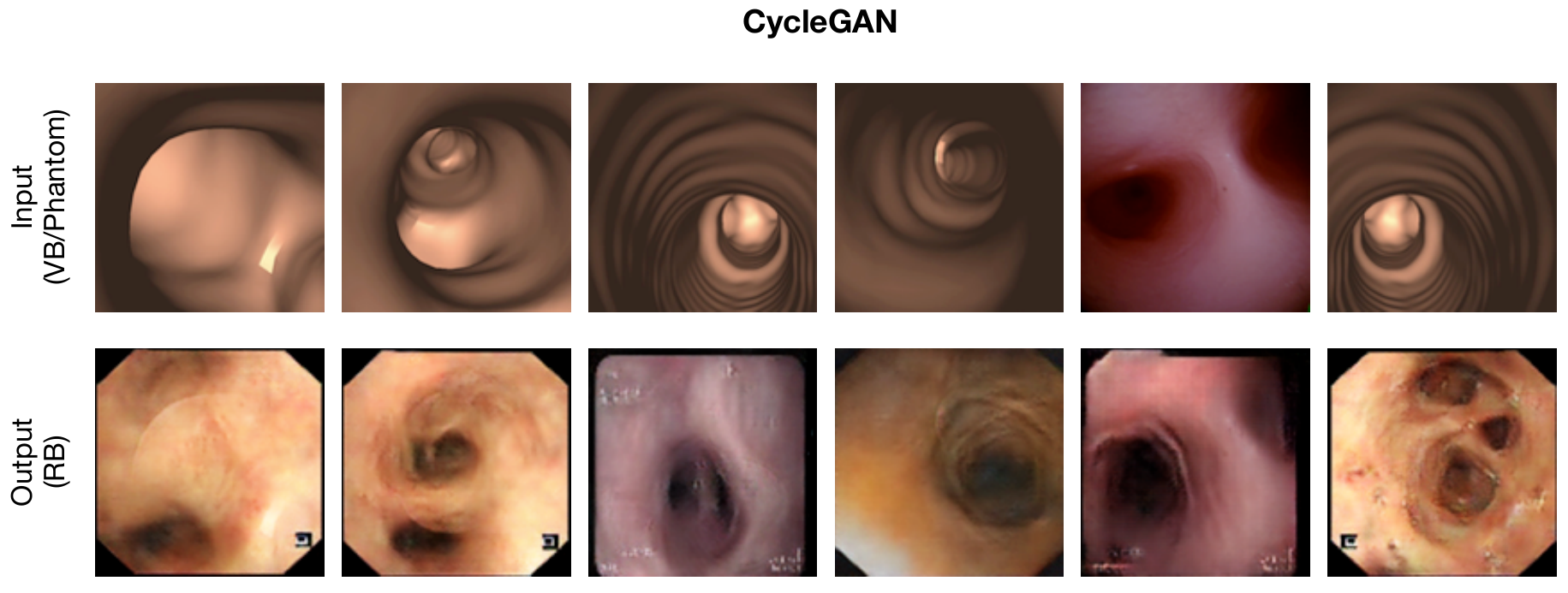}
    \caption{Qualitative results for baseline model CycleGAN. Input was an RGB image from VB and phantom datasets. Note that bronchial orifices are not preserved in multiple cases.}
    \label{fig:cyclegan_results}
\end{figure*}

\begin{figure*}[h]
    \centering
    \includegraphics[width=.99\textwidth]{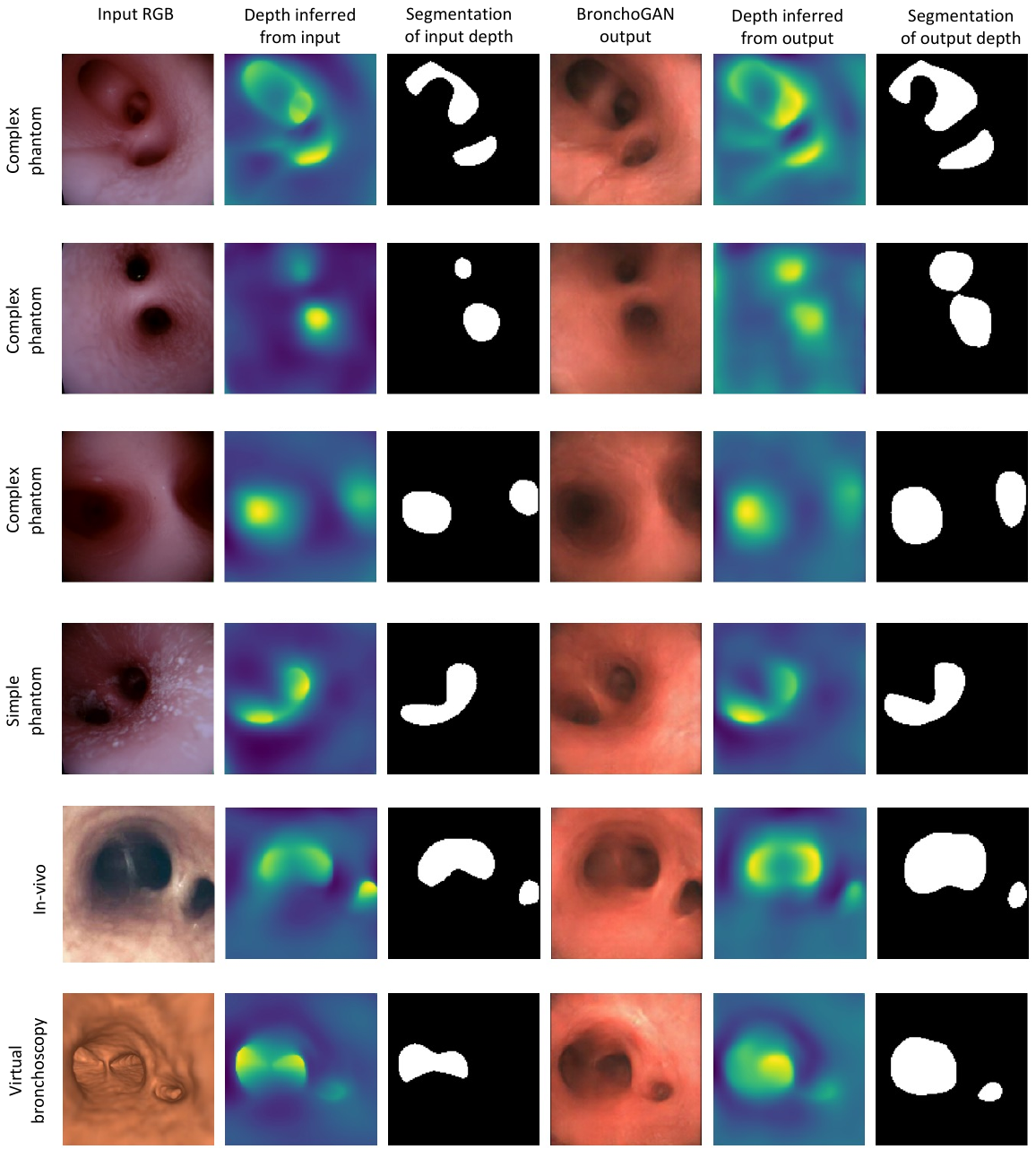}
    \caption{Results obtained using our method \textbf{BronchoGAN}. It shows input RGB images (VB, RB, phantom), depth images inferred from input RGB images using depthAnything and extracted bronchial orifices thereof. Depth images were inferred again from the generated (GAN) output image. 
    \textbf{Anatomical constraints:} Penalization depending on segmentation maps constructed from input's and output's depth images (dice loss).}
    \label{fig:bronchogan_results}
\end{figure*}

\begin{figure*}[h]
    \centering
    \includegraphics[width=.99\textwidth]{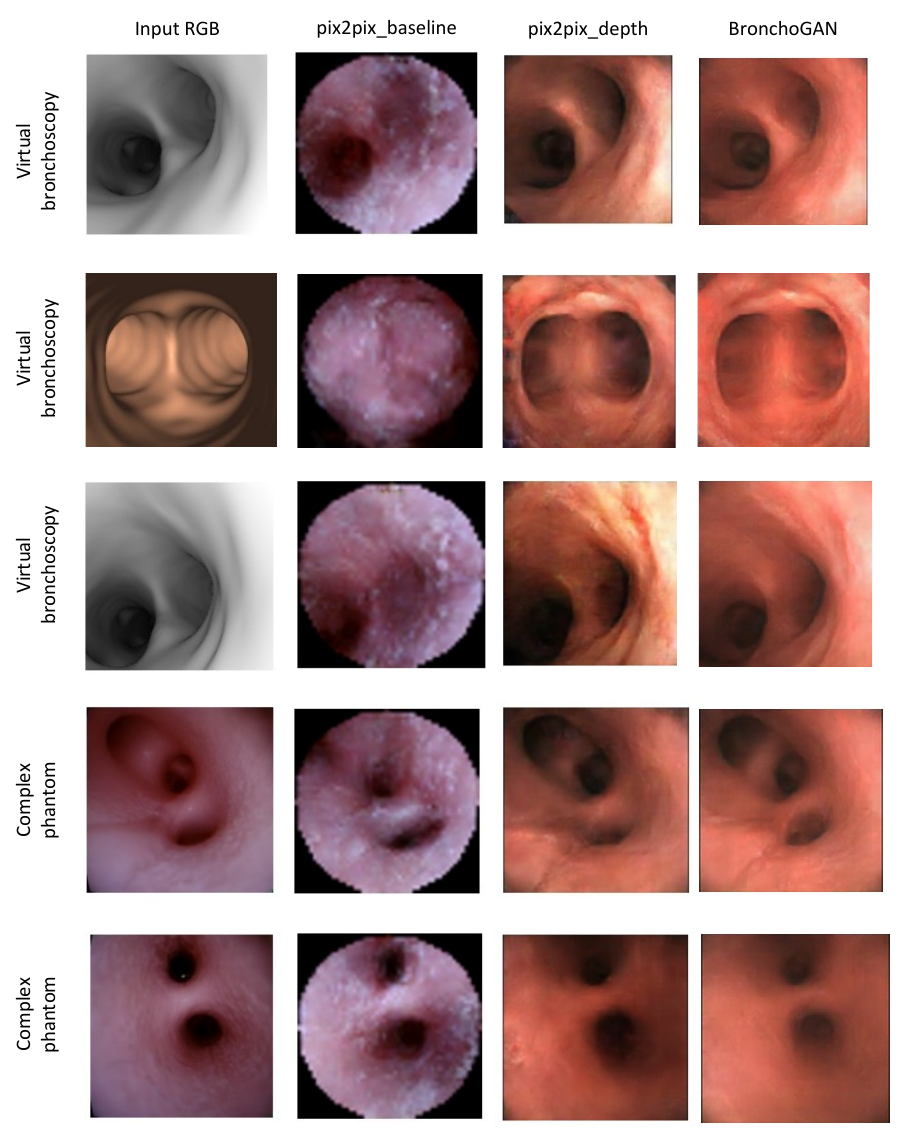}
    \caption{Qualitative comparison of our proposed models BronchoGAN and pix2pix\_depth vs. the baseline pix2pix\_base. Note that pix2pix\_base was unable to preserve input bronchial orifices in multiple cases. Also, domain gaps become apparent in row 2.}
    \label{fig:all_gan_results}
\end{figure*}

Fig. \ref{fig:bronchogan_results} illustrates resulting outputs and intermediate steps for our proposed models, pix2pix\_depth and BronchoGAN. Fig.  \ref{fig:cyclegan_results}, \ref{fig:bronchogan_results} and \ref{fig:all_gan_results} show visual outputs of all investigated GAN models. Quantitative results are shown in Table \ref{tab:evaluation_metrics}.

\subsection{Results}

\begin{table}[ht]
\centering
\begin{tabular}{l|ccccc}
\hline
Model & FID ↓ & SSIM ↑ & DICE ↑ \\
\hline
cycleGAN & 1717.9574 & 0.2831 & 0.2412 \\
pix2pix\_base & 1564.0430 & 0.4042 & 0.3950 \\
pix2pix\_depth (ours) & 1006.5910 & 0.3875 & 0.6334 \\
BronchoGAN (ours) & \textbf{770.6833} & \textbf{0.4623} & \textbf{0.6743} \\

\hline
\end{tabular}
\caption{Quantitative results obtained for $2\,271$ VB test images of the Harvard image dataset \cite{banach2022_harvard_data}. Dice coefficients were estimated based on input and synthesized image bronchial orifice segmentations obtained with our training-free pipeline. \cite{keuth_weakly_2022}.}
\label{tab:evaluation_metrics}
\end{table}

\subsection*{Discussion}
Our experiments highlight significant advantages of leveraging depth images (inferred from a foundation model) as intermediate representations for image translation tasks, as opposed to directly training on RGB images. This approach not only facilitates more robust translation results but also simplifies the process of constructing image pairs for training. When comparing visual outputs and quantitative metrics of BronchoGAN and pix2pix\_depth, the differences are less evident than when contrasted with the baselines. Slightly better dice coefficients for BronchoGAN demonstrated improved preservation of bronchial orifices. Depth images inherently encode airway entrances, as they predominantly capture the distal structures within the bronchoscopic field of view. We expect that this is the reason why we could not observe more significant differences when comparing our approaches pix2pix\_depth and BronchoGAN. However, we anticipate that the benefit of incorporating anatomical constraints will become more obvious when incorporating patients cohorts with $> 1\,000$ patients as airway abnormalities are not as frequent in the population as already reported in \cite{keuth2024airway}. Also, when a substantial larger anatomical variation from VB data faces a limited amount of RB images (with limited anatomical variation) we expect more significant differences with our GAN being strongly guided by the segmentation priors.   

\noindent \textbf{Limitations}. The requirement of circularly cropped images for pix2pix\_base slightly limits the (visual) comparison to the other models (our quantitative comparison take this into consideration). However, this is necessary since aligned image pairs were constructed in this way \cite{visentini-scarzanella_deep_2017}. The anatomical variance covered by our training and test data is currently limited. 

\section{Conclusion}\label{sec:conclusion}
We proposed two novelties in the context of GAN-based image translation for bronchoscopy: (a) an intermediate image representation spanned by a robust depth foundation model closing frequently reported domain gaps in bronchoscopy due to limited data availability \cite{keuth_weakly_2022,keuth2024airway} and (b) the utilization of anatomical guidance penalizing inadequate overlap of bronchial orifices measured on input and output images. We showed that both elements contribute to domain-agnostic and anatomically stable image translation in bronchoscopy. In our future work, we will incorporate VB images from large-scale public CT datasets to further stabilize translation and to enable more exhaustive testing of varying anatomies. BronchoGAN can substantially contribute to training more robust models for bronchoscopy image synthesis closing domain gaps due to limited datasets. This will become evident for promoting deep learning-based visual guidance for interventions in pulmonology.

\section*{Declarations}
The authors have no relevant financial or non-financial interests to disclose.

\clearpage
\bibliography{references}


\end{document}